# An integrated cryogenic optical modulator

Authors: Felix Eltes[1], Gerardo E. Villarreal-Garcia[2], Daniele Caimi[1], Heinz Siegwart[1], Antonio A. Gentile[2], Andy Hart[2], Pascal Stark[1], Graham D. Marshall[2], Mark G. Thompson[2], Jorge Barreto[2], Jean Fompeyrine[1], Stefan Abel[1]

[1] IBM Research – Zurich, Säumerstrasse 4, 8803 Rüschlikon, Switzerland

[2] Quantum Engineering Technology Labs, H. H. Wills Physics Laboratory, University of Bristol, Bristol, BS8 1TL, United Kingdom

**Integrated electrical and photonic circuits (PIC) operating at cryogenic temperatures are fundamental building blocks required to achieve scalable quantum computing, and cryogenic computing technologies[1–4]. Optical interconnects offer better performance and thermal insulation than electrical wires and are imperative for true quantum communication. Silicon PICs have matured for room temperature applications but their cryogenic performance is limited by the absence of efficient low temperature electro-optic (EO) modulation. While detectors and lasers perform better at low temperature[5,6], cryogenic optical switching remains an unsolved challenge. Here we demonstrate EO switching and modulation from room temperature down to 4 K by using the Pockels effect in integrated barium titanate ($BaTiO_3$)-based devices[7]. We report the nonlinear optical (NLO) properties of $BaTiO_3$ in a temperature range which has previously not been explored, showing an effective Pockels coefficient of 200 pm/V at 4 K. We demonstrate the largest EO bandwidth (30 GHz) of any cryogenic switch to date, ultra-low-power tuning which is $10^9$ times more efficient than thermal tuning, and high-speed data modulation at 20 Gbps. Our results demonstrate a missing component for cryogenic PICs. It removes major roadblocks for the realisation of novel cryogenic-compatible systems in the field of quantum computing and supercomputing, and for interfacing those systems with the real world at room-temperature.**

Cryogenic technologies are becoming essential for future computing systems, a trend fuelled by the world-wide quest to develop quantum computing systems and future generations of high-performance classical computing systems[8,9]. While most computing architectures rely solely on electronic circuits, photonic components are becoming increasingly important in two areas. First, PICs can be used for quantum computing approaches where the quantum nature of photons is exploited as qubits[3,4]. Second, optical interconnects can overcome limitations in bandwidth and heat leakage that are present in conventional electrical interconnect solutions for digital data transfer between cryogenic processors and the room temperature environment[2]. In addition, due to their low interaction with the environment, photons are the only viable carriers to transport quantum states over large distances. Optical interfaces are therefore essential for true quantum communication, necessary to connect multiple quantum computers[10,11] and for secure remote operation of quantum computers[12].

Today, the realisation of such photonic concepts is hindered by the lack of switches and modulators that operate at cryogenic temperatures with low-loss, high bandwidth, and low static power consumption. So far, only two concepts for cryogenic EO switches have been investigated, based either on the thermo-optic effect[13] or the plasma-dispersion effect[14]. Both mechanisms have physical limitations which intrinsically restrict the low-temperature performance of such devices. Thermo-optic phase shifters exploit Joule heating with large static power consumption and exhibit a bandwidth of less than a few MHz[15]. Plasma-dispersion-based devices require very high doping levels to compensate for carrier freeze-out at cryogenic temperatures. The high doping leads to large propagation losses and devices are limited to a bandwidth of <5 GHz in micro-disk modulators[14]. The use of EO switches based on the Pockels effect has been shown to offer low propagation losses and high-bandwidth, combined with low static power consumption at room temperature[7,16–18]. Because of its electro-static nature, the Pockels effect has no intrinsic physical limitations for its application at cryogenic temperature. However, the lack of a Pockels effect in silicon means that heterogenous integration of new materials is needed to bring Pockels-based switching in PIC platforms. Pockels modulators have recently been demonstrated using organics[19], PZT[16], LiNbO$_3$[20], and BaTiO$_3$[7]. Among them, BaTiO$_3$ stands out due to having the largest Pockels coefficients[7] and exhibiting compatibility with advanced silicon photonics platforms[21]. We complete this triumvirate by demonstrating that BaTiO$_3$ is also an ideal candidate for cryogenic EO integration.

Both the NLO properties and structural behaviour of BaTiO$_3$ thin-films are entirely unknown at temperatures below 300 K. In fact, even in bulk BaTiO$_3$ crystals the NLO behaviour is unexplored below 270 K, and the room temperature NLO behaviour of BaTiO$_3$ thin-films has only recently been thoroughly investigated[7]. Predictions of the Pockels tensor at cryogenic temperatures based on data at higher temperatures is not possible because the temperature dependence of neither the Pockels coefficients nor the crystalline phase of thin-film $BaTiO_3$ on Si is known. The phase transitions of thin-films are expected to differ from bulk crystals[22] due to the structural mismatch and thermal stress that exists between the substrate and the $BaTiO_3$ layer[23–25]. Here, we determine the cryogenic behaviour of $BaTiO_3$ thin films by analysing the performance of $BaTiO_3$-based EO switches at temperatures down to 4 K. Our results show that efficient EO switching at cryogenic temperature is indeed possible and with bandwidths beyond 30 GHz. We also demonstrate the applicability of such devices for low-power switching and tuning as well as high-speed data modulation at 20 Gbps at 4 K.

In this work, we use two waveguide designs fabricated on single crystalline $BaTiO_3$ layers bonded to $SiO_2$-buffered silicon substrates (Figure 1a, see Methods). In the first design, silicon nitride (SiN)-based waveguides allowed us to study the pure NLO properties of $BaTiO_3$ in absence of mobile charge carriers which could result in an additional, non-Pockels EO response. In the second, silicon (Si) waveguides served as more efficient devices to demonstrate high-speed data modulation. The enhanced efficiency originates from a larger optical-mode overlap with the $BaTiO_3$ layer (41 %) than with the SiN waveguides (18 %) (Figure 1b,c). We found that the propagation losses (5.6 dB/cm, SiN device) were not affected by the presence of $BaTiO_3$ in the active section (Supplementary Note, SN 1) throughout the temperature range studied.

To characterise the NLO behaviour of $BaTiO_3$ at 4 K, we measured the induced resonance shift in a racetrack resonator as a function of the DC bias (Figure 1d,e), from which we determined the refractive index change of $BaTiO_3$ ($\Delta n_{BTO}$) as a function of the applied electric field (see Methods). This dependence allows us to study two of the three expected features of Pockels-based switching[7]: NLO hysteresis and angular anisotropy, the third being the persistence of the Pockels effect at high frequencies (>10 GHz)[26].

The NLO response with a hysteretic behaviour (Figure 2a) indicates that a non-vanishing Pockels effect is preserved in $BaTiO_3$ down to a temperature of 4 K. We determine the effective Pockels coefficient, $r_{eff}$, by analysing the hysteretic behaviour of the refractive index change (SN 2). The dependence of $r_{eff}$ on device orientation (Figure 2b) reveals the second signature of the Pockels effect, its anisotropy. The reduced magnitude at 4 K compared to room temperature is due to a temperature dependence of the Pockels effect, as discussed below. While $r_{eff}$ is reduced with temperature, the EO response is expected to be present at high frequencies also at low temperature. Indeed, we observe a constant EO response in racetrack resonators with a low $Q$ factor ($Q \sim 1{,}800$) up to 30 GHz (Figure 2b). This constitutes the highest bandwidth for any cryogenic modulator reported to date. The frequency response is expected to remain flat at even higher frequency but could not be measured in our experiment (see Methods). The hysteretic behaviour, anisotropy, and high-speed response prove the presence of the Pockels effect in $BaTiO_3$ at 4 K.

We performed electrical characterisation of the material at low temperature using dedicated electrical test structures (SN 3). The resistivity of $BaTiO_3$ at 4 K is very high, $>10^9$ $\Omega$m. In fact, the measured current is dominated by capacitive charging and ferroelectric switching currents (Figure 2d). The field-dependent capacitance shows clear hysteretic characteristics (Figure 2e), consistent with ferroelectric domain switching.

The measured $r_{eff}$ at 4 K is lower than at room temperature (Figure 2b), which has two causes. First, the Pockels effect itself is generally temperature dependent due to changes in strain and polarisation of the crystal[27]. Second, the non-zero elements of the Pockels tensor depend on the crystal symmetry, which can change abruptly with temperature due to structural phase transitions. $BaTiO_3$ bulk crystals are known to transition from a tetragonal phase at room temperature to orthorhombic and rhombohedral phases at lower temperatures (~270 K and ~200 K respectively)[28]. Such transitions change the elements of the Pockels tensor and modify the magnitude of the effective Pockels coefficients[27]. Because phase transitions of thin-film materials can be drastically affected by substrate strain[23–25], studying the properties of thin-film $BaTiO_3$ becomes critical when considering cryogenic applications. To investigate the effects of possible phase transitions, we measured $r_{eff}$ in a range from 4 to 340 K. Indeed, the magnitude of $r_{eff}$ is strongly temperature-dependent (Figure 3). A peak around 240 K, with $r_{eff} >700$ pm/V, is consistent with the reported divergence of the $r_{42}$ element of the Pockels tensor close to the tetragonal-orthorhombic transition[27]. Consistently, the permittivity of the $BaTiO_3$ layer (see Methods) also shows a peak in the same temperature range (SN 4), confirming that the abrupt change in $r_{eff}$ is caused by a phase transition. Below 240 K the magnitude of $r_{eff}$ decreases gradually to around 200 pm/V at 4 K. In addition to the phase transition at 240 K, a second phase transition occurs below 100 K causing a rapid change in $r_{eff}$ of 90° devices. This phase transition is also observed in the qualitative behaviour of the NLO hysteresis which shows that the transitions is induced by the electric field (SN 4). While $r_{eff}$ of BaTiO$_3$ is reduced at 4 K compared to room temperature, the value of ~200 pm/V is still larger than most other material systems at room temperature[16,29,30]. The effect of a reduced Pockels coefficient on the energy efficiency of EO switching is partially compensated for by a simultaneous reduction of the permittivity of BaTiO$_3$ (SN 4). Additionally, the conductivity of BaTiO$_3$ is reduced by more than four orders of magnitude (SN 3), resulting in a negligible static power consumption of BaTiO$_3$-devices in cryogenic environments.

We demonstrate the applicability of BaTiO$_3$ for cryogenic photonic applications by two examples: low-power EO switching and high-speed data modulation. For switching we use a Mach-Zehnder interferometer with 2×2 multimode interference splitters, applying a voltage to one arm. Because the leakage current through BaTiO$_3$ at 4 K is 10$^4$ times lower than at 300 K, less than 10 pW static power is consumed when inducing a π phase shift to switch between the two optical outputs (Figure 4a,b). Compared to state-of-the-art technology based on thermo-optic phase shifters[13], static tuning using BaTiO$_3$ is one billion times more power efficient. The dynamic energy of the switch is ~30 pJ, which could be reduced to ~2 pJ in an optimised device geometry (SN 5). As a second example, we performed data modulation experiments by sending a pseudo-random bit-sequence to the modulator and recording the optical eye-diagram (Figure 4c,d). Data transmission at rates up to 20 Gbps are achieved with our experimental setup using a drive voltage ($V_{pp}$) of just 1.7 V, resulting in an extremely low energy consumption of 45 fJ/bit.

In conclusion, we have shown that BaTiO$_3$ thin films can be used to realise EO switches and modulators for efficient cryogenic operation of silicon PICs. We have demonstrated low-power switching, as well as high-speed data modulation. Combining BaTiO$_3$ with silicon photonic integrated circuits, we make a building block available that was previously inaccessible for any cryogenic circuits. We anticipate that such new components are a milestone for a versatile platform of cryogenic photonics for applications such as quantum computing and cryogenic computing technology, as well as quantum interconnects to room-temperature environments.

# Acknowledgements

This work has received funding from the European Commission under grant agreements no. H2020-ICT-2015-25- 688579 (PHRESCO) and H2020-ICT-2017-1-780997 (plaCMOS), from the Swiss State Secretariat for Education, Research and Innovation under contract no. 15.0285 and 16.0001, from the Swiss National Foundation project no 200021_159565 PADOMO, from EPSRC grants EP/L024020/1, EP/M013472/1, and EP/K033085/1, the UK EPSRC grant QuPIC (EP/N015126/1), and ERC grant 2014- STG 640079. JB thanks Dr. Döndü Sahin for her assistance with the experimental setup.

# Methods

**Device fabrication.** Single crystalline BaTiO$_3$ was deposited on top of an epitaxial 4-nm-thick STO seed layer by molecular beam epitaxy on 8" SOI wafers with 220-nm-thick device silicon layer for SiN-based devices, and on 2" SOI wafers with 100-nm-thick device silicon for Si-based devices, following a process described elsewhere[7]. Direct wafer bonding was used to transfer the BaTiO$_3$ and device Si layers onto high-resistivity Si wafers capped with a 3-µm-thick thermal oxide. Specifics of the direct wafer bonding process can be found in ref. 7.

For SiN-based waveguides the device Si layer was removed by dry etching, followed by chemical vapor deposition of SiN. The waveguide layer (Si or SiN) was patterned by dry etching. After waveguide patterning, a combination of SiO$_2$ cladding deposition, via etching, and metallisation was used to form the final cross-section. Intermediate annealing steps were used to reduce propagation losses.

The SiN-based waveguides use an 80-nm-thick BaTiO$_3$ layer, and 150-nm-thick SiN layer. The strip-waveguide width is 1.1 µm. The electrode-to-electrode gap is 9 µm. The Si strip-waveguides were fabricated using 225-nm-thick BaTiO$_3$ and 100-nm-thick Si. The waveguide width is 0.75 µm, and the electrode-to-electrode gap is 2.3 µm.

**Cryogenic measurements.** The cryogenic electro-optic measurements were performed in a Lakeshore CPX cryogenic probe station, fitted with RF (40 GHz BW, K-type connectors) and optical feed-throughs. DC and RF signals were applied to the devices using RF probes, and optical coupling was achieved using a fibre array with polarisation maintaining fibres for 1550 nm. A tuneable laser (EXFO T100S-HP) and power meter were used to record transmission spectra (EXFO CT440). The cryogenic electrical measurements were performed in a Janis cryogenic probe station equipped with DC probes. Current-voltage and capacitance-voltage measurements were performed using a parameter analyser. Both cryogenic probe stations were cooled by liquid helium to a base temperature of 4.2 K.

**DC EO characterisation.** The DC electro-optic response was extracted by applying a voltage to the electrodes of a racetrack resonator and measuring the shift in resonance wavelength ($\Delta\lambda$), compared to the unbiased case, as a function of the applied voltage. From the measured wavelength shift, the change in BaTiO$_3$ refractive index ($\Delta n_{\mathrm{BTO}}$) can be estimated as

$$\Delta n_{\text{BTO}} = \frac{\lambda_0 \cdot \Delta \lambda}{FSR \cdot L_E \cdot \Gamma_{\text{BTO}}}$$

where $\Gamma_{\text{BTO}}$ is the optical confinement in BaTiO$_3$, FSR is the free spectral range of the resonator, $L_E$ is the electrode length, and $\lambda_0$ is the resonance wavelength with no voltage.[7] The effective Pockels coefficient, $r_{\text{eff}}$, was then determined according to the procedure described in SN 2.

**RF frequency response.** To measure the EO frequency response (EO $S_{21}$) a vector network analyser (VNA, Keysight PNA 50 GHz) was used to apply the electrical stimulus to a BaTiO$_3$ ring modulator. The modulated optical signal was applied to a photodiode (Newport 1024) and the response recorded by the VNA. Electrical calibration was performed before the measurement, and the response of the photodetector was compensated for the data analysis. While the VNA could generate signals up to 50 GHz, the bandwidth of the photodetector was 26 GHz, which in combination with large frequency-dependent electrical losses in the cryogenic probe station (SN 6) makes it impossible to measure the bandwidth beyond 30 GHz.

**Devices for data modulation.** For data modulation experiments, devices with Si strip-waveguide were used. The racetrack resonator that was used had a bending radius of 15 µm and straight sections of 30 µm.

**Data modulation experiments.** The electrical signal was generated using an arbitrary waveform generator. A pseudo-random bit stream of $2^7-1$ bits was used for modulation. The electrical signal was pre-distorted to compensate for the finite time-response of the electrical signal path (SN 6). The signal was amplified using a RF amplifier and sent to the cryogenic setup, with an estimated voltage swing on the device of 1.7 V (SN 6). A Pritel FA-23 EDFA was used to amplify the modulated optical signal which was applied to a photo diode and recorded on an oscilloscope.

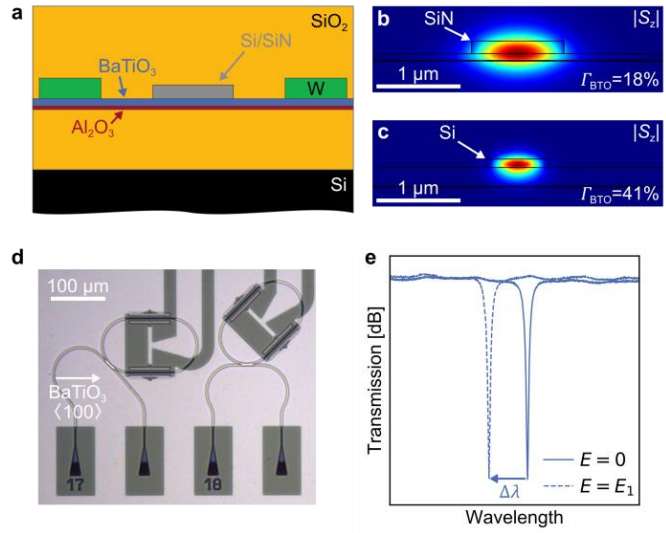

**Figure 1. BaTiO₃ electro-optic device concept. a**, Schematic cross-section of the devices. A silicon or silicon nitride layer forms a strip-waveguide on top of an BaTiO₃ layer. Lateral electrodes fabricated with W are used to apply an electric field across the BaTiO₃. The devices are embedded in SiO₂ layers on top of silicon substrates. **b**, Optical mode simulation of the SiN waveguide geometry and **c** the Si waveguide geometry, showing an optical confinement in BaTiO₃ of 18 % and 41 % respectively. **d**, Optical micrograph of racetrack resonator devices used to characterise the nonlinear optical properties of BaTiO₃. **e**, Characterisation principle of resonant electro-optic switches, showing example data of a shifted resonance. The shift in resonance wavelength is measured for an applied electric field and converted to the material properties of BaTiO₃ (see Methods).

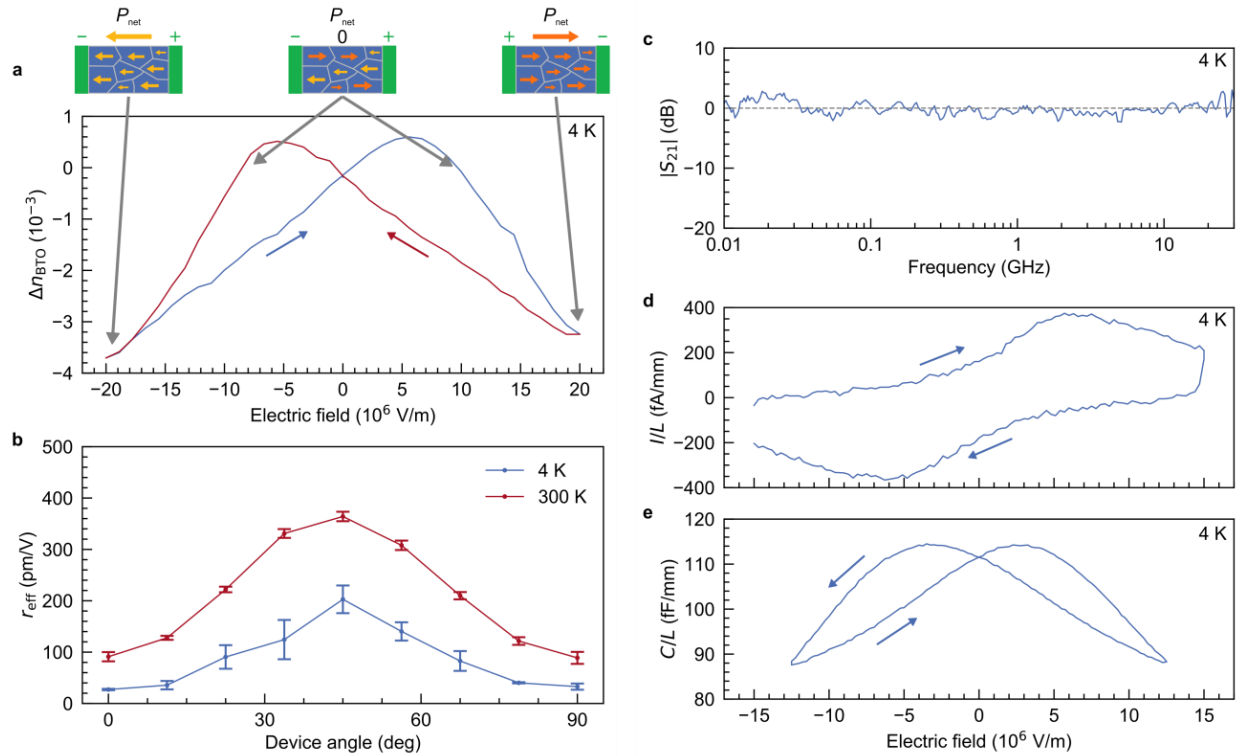

**Figure 2. Electro-optic and electrical response of BaTiO$_3$-based optical switches at 4 K. a**, Refractive index change of BaTiO$_3$ as a function of applied electric field for a device in the 11.25° direction (as defined in b). The hysteretic behaviour originates from ferroelectric domain switching in the BaTiO$_3$, as shown schematically (top). **b**, Angular anisotropy of the effective Pockels coefficient in BaTiO$_3$. The angle is defined relative to the BaTiO$_3$<100> direction. The same anisotropy as for BaTiO$_3$ at room temperature is observed but with reduced magnitude. The error bars show the combined standard error of the fit and from averaging measurements of multiple devices with the same orientation. **c**, Electro-optic $S_{21}$-parameter of BaTiO$_3$ ring resonator showing a flat response up to a record frequency of 30 GHz at 4 K. **d**, Current measured as a function of electric field across the BaTiO$_3$ layer showing extremely low current flowing through the material. The current is dominated by capacitive charging together with ferroelectric switching current resulting in a peak (SN 3). **e**, Capacitance as a function of electric field, showing characteristic ferroelectric hysteresis and field-dependent permittivity.

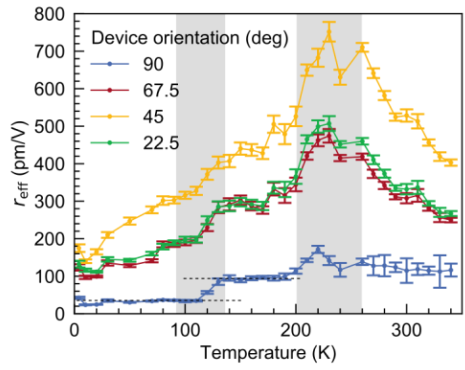

**Figure 3. Temperature dependence of the Pockels effect in BaTiO$_3$.** The effective Pockels coefficient along different crystal orientations at temperatures from 4 K to 340 K. The peak around 240 K is the signature of a phase transition in BaTiO$_3$. A second, field-induced phase transition occurs around 100 K, causing a sharp drop of $r_{eff}$ in 90° devices (indicated by horizontal dashed lines). This phase transition is also evident in the qualitative evaluation of the optical response (SN 4). The grey areas indicate the temperature ranges of the respective phase transitions. The error bars show the standard error of the fit used to extract the Pockels coefficients (SN 2).

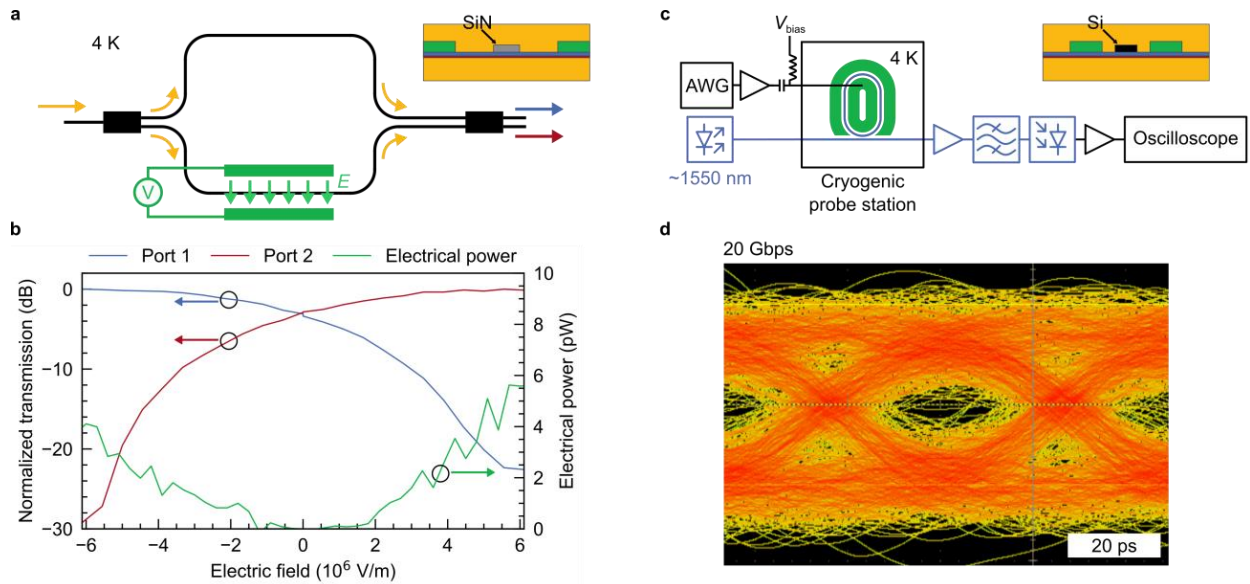

**Figure 4. Demonstration of low-power switching and high-speed data modulation with BaTiO$_3$-based devices at 4 K. a**, Schematic of Mach-Zehnder (MZ) configuration used to switch between two ports. The inset shows the waveguide cross-section. **b**, Transmission from both ports of a MZ switch as a function of applied electric field, along with the static power consumption. When fully switching between outputs, less than 10 pW static power is consumed, and only 30 pJ of dynamic energy. **c**, Schematic of the experimental setup for data modulation. The electrical signal was amplified to compensate for losses into the cryogenic probe station, and then applied to the device. The modulated optical signal was detected using a photodiode and recorded on a high-speed oscilloscope. The inset show the waveguide cross-section. **d**, Eye diagram recorded at 20 Gbps with $V_{pp}$ = 1.7 V, corresponding to modulation energy of 45 fJ/bit.